\documentclass[10pt,conference]{IEEEtran}

\IEEEoverridecommandlockouts

\usepackage{cite}

\usepackage[cmex10]{amsmath}
\usepackage{array}
\ifCLASSINFOpdf
   \usepackage[pdftex]{graphicx}
   \DeclareGraphicsExtensions{.pdf,.jpeg,.png}
\else
   \usepackage[dvips]{graphicx}
   \graphicspath{{../eps/}}
   \DeclareGraphicsExtensions{.eps}
\fi

\begin{document}

\title{Algebraic Attack on the Alternating Step($r$,$s$) Generator}


%

\author{ \IEEEauthorblockN{Mehdi M. Hassanzadeh}
\thanks{This work was supported by the Norwegian Research
Council and partially by the grant NIL-I-004 from Iceland,
Liechtenstein and Norway through the EEA and Norwegian Financial
Mechanisms.}
				 \IEEEauthorblockA{The Selmer Center\\
												Department of Informatics, University of Bergen\\
												P.O. Box 7800, N-5020 Bergen, Norway\\
												Email: Mehdi.hassanzadeh@ii.uib.no}\and
				 \IEEEauthorblockN{Tor Helleseth}
				 \IEEEauthorblockA{The Selmer Center\\
												Department of Informatics, University of Bergen\\
												P.O. Box 7800, N-5020 Bergen, Norway\\
												Email: Tor.helleseth@ii.uib.no}
											
			}

\maketitle

\begin{abstract}

The Alternating Step$(r,s)$ Generator, ASG$(r,s)$, is a clock-controlled sequence generator which is recently proposed by A. Kanso. It consists of three registers of length $l$, $m$ and $n$ bits. The first register controls the clocking of the two others.  The two other registers are clocked $r$ times (or not clocked) (resp. $s$ times or not clocked) depending on the clock-control bit in the first register. The special case $r=s=1$ is the original and well known Alternating Step Generator. Kanso claims there is no efficient attack against the ASG$(r,s)$ since $r$ and $s$ are kept secret. In this paper, we present an Alternating Step Generator, ASG, model for the ASG$(r,s)$ and also we present a new and efficient algebraic attack on ASG$(r,s)$ using $3(m+n)$ bits of the output sequence to find the secret key with $O((m^2+n^2)2^{l+1}+ m^32^{m-1} + n^32^{n-1})$ computational complexity. We show that this system is no more secure than the original ASG, in contrast to the claim of the ASG$(r,s)$'s constructor. 
\end{abstract}

\section{Introduction}
The goal in stream cipher design is to efficiently produce pseudorandom sequences which should be indistinguishable from truly random sequences. 
From a cryptanalysis point of view, a good stream cipher should be resistant against a \textit{known-plaintext attack}. In this kind of attack, the cryptanalyst is given a plaintext and the corresponding ciphertext, and the task is to determine the secret key. For a synchronous stream cipher, this is equivalent to the problem of finding the secret key or initial state that produced a given keystream output.

In stream cipher design, one usually uses Linear Feedback Shift Registers, LFSRs, as building block in different ways, and the secret key is often used as the initial state of the LFSRs. A general methodology for producing random-like sequences from LFSRs that has been popular is using the output of one or more LFSRs to control the clock of other LFSRs. The purpose is to destroy the linearity of the LFSR sequences and hence provide the resulting sequence with a large linear complexity. This structure is called a \textit{Clock-Controlled Generator} which has several different types, e.g., Stop/Go Generator \cite{R[2], R[3]}, Step1/Step2 Generator \cite{R[3]}, Shrinking Generator \cite{R[4]}, Self-Shrinking Generator \cite{R[5]}, and Jump Register which is proposed recently in \cite{R[6], R[7], R[8]} and it is used in some candidates to the European ECRYPT/eSTREAM \cite{R[9]} project, e.g., Pomaranch \cite{R[10]} and Mickey \cite{R[11]}.

An Alternating Step Generator (ASG), a well-known stream cipher proposed in \cite{R[12]}, consists of a regularly clocked binary LFSR, \textbf{A}, and two Stop/Go clocked binary LFSRs, \textbf{B} and \textbf{C}. At each time, the clock-control bit from \textbf{A} determines which one of the two Stop/Go LFSRs is clocked, and the output sequence is obtained as bit-wise sum of the two Stop/Go clocked LFSR sequences.

ASG$(r,s)$ proposed in \cite{R[1]} is a general form of an original ASG which will be described in the next section. The difference is that \textbf{B} and \textbf{C} are shifted $r$ and $s$ times, respectively, where $r$ and $s$ are part of the secret key. As far as we know, there is presently no efficient general attack on this algorithm. In this paper, we propose an algebraic attack on this algorithm and we will show that its security is no more than the security of the original ASG, in contrast to the constructor's claim.

In Section II, a brief description of the ASG$(r,s)$ will be presented and in Section III, the security of the ASG$(r,s)$ is investigated from the author's point of view. We model the ASG$(r,s)$ to an original ASG in Section IV and according to this model, we will present our attack in Section V and conclude in Section VI.

\section{Description of the ASG$(r,s)$}
The \textbf{A}lternating \textbf{S}tep$(r,s)$ \textbf{G}enerator, ASG$(r,s)$, is a clock-controlled based stream cipher and it is similar to the original ASG but the clock-controlled LFSR \textbf{B} and \textbf{C} jump $r$ and $s$ steps respectively instead of in a Stop/Go manner.

ASG$(r,s)$ is composed of a regularly clocked FSR, \textbf{A}, and two clock-controlled FSR's, \textbf{B} and \textbf{C}. At each time, the clock-control bit from \textbf{A}, e.g., $0^{th}$ cell, determines which of the two FSR's is clocked. \textbf{B} is clocked by the constant integer $r$ and \textbf{C} is not clocked if the content of the $0^{th}$ cell of \textbf{A} is `1', otherwise, \textbf{B} is not clocked and \textbf{C} is clocked by the constant integer $s$. FSR \textbf{A} is called the Control Register and FSRs \textbf{B} and \textbf{C} are called the Generating Registers. The output bits of the ASG$(r,s)$ are produced by adding modulo 2 the output bits of FSRs \textbf{B} and \textbf{C} under the control of FSR \textbf{A}.
\begin{figure}[!t]
	\centering
		\includegraphics[width=3.5in]{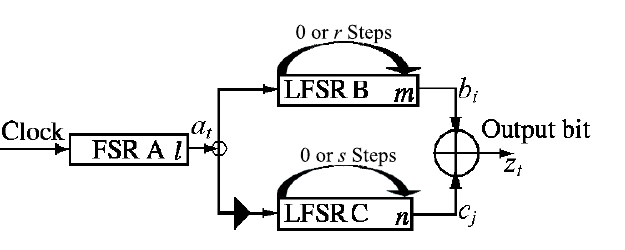}
	\caption{The Alternating Step$(r,s)$ Generator, ASG$(r,s)$}
	\label{fig1}
\end{figure}
Kanso has recommended using a FSR \textbf{A} with a de-Bruijn output sequence of span $l$ \cite{R[14]} and Primitive Linear Feedback Shift Register (LFSR) for generating registers \textbf{B} and \textbf{C} with length $m$ and $n$ bits respectively which is illustrated in fig. \ref{fig1}. He proved that when the values of $m$ and $n$ are satisfying $\gcd(m,n)=1$, and the values of $r$ and $s$ are satisfying $\gcd(r,2^m-1)=1$ and $\gcd(s, 2^n-1)=1$, then the period of the output sequences is equal to $2^l(2^m-1)(2^n-1)$ and the linear complexity  ($L_l$) of the output sequence satisfies $(m+n)2^{l-1} < L_l \leq (m+n)2^l$. The initial states of registers and the number of jumps, $r$ and $s$, are the secret key. This structure is considered in the whole paper and in our attack.

\section{SECURITY OF THE ASG$(r,s)$}
Kanso claims in \cite{R[1]} that his structure, ASG$(r,s)$, is secure against all known attacks so far. The output sequence of the ASG$(r,s)$ is the XOR of its two irregularly decimated generating sequences. Thus, he claims one could not expect a strong correlation to be obtained efficiently, especially, if the primitive feedback polynomials of high Hamming weight are associated with the feedback functions of the generating registers \textbf{B} and \textbf{C} \cite{R[23]}. Furthermore, the values of $r$ and $s$ are considered as part of the secret key. Then, ASG$(r,s)$ appears to be secure against all correlation attacks introduced in \cite{R[20], R[23], R[24],R[25],R[26],R[27],R[28],R[29],R[30],R[31]}.

Kanso also made the claim that ASG$(r,s)$ is secure against  algebraic attacks \cite{R[13]} and the complexity of this attack is equal to $O((m^3+n^3)\Phi 2^l)$\footnote{In \cite{R[1]}, it is mentioned that this complexity is $O(\Phi 2^l m^3 n^3)$ which is not correct.}  where $\Phi=\Phi_1 \Phi_2$, $\Phi_1$ is the number of possible values for $r$ such that $\gcd(r, 2^m-1)=1$ and $\Phi_2$ is the number of possible values for $s$ such that $\gcd(s, 2^n-1)=1$. This attack takes approximately $O((m^3+n^3)2^{m+n+l-2})$ steps using the estimate $\Phi_1=2^{m-1}$ and $\Phi_2=2^{n-1}$. Therefore, the ASG$(r,s)$ appears to be secure against this attack.

\section{ASG MODEL FOR THE ASG$(r,s)$}
Throughout the paper, we refer to the output sequence of registers \textbf{A}, \textbf{B} and \textbf{C} by $a= a_0, a_1, ..., a_t$, $b= b_0, b_1, ..., b_i$ and $c= c_0, c_1, ..., c_j$ respectively. Furthermore, we refer to the output sequence of the ASG$(r,s)$ by $z= z_0, z_1, ..., z_t$. Let $S_a(t)$, $S_b(t)$ and $S_c(t)$ denote the internal states of registers \textbf{A}, \textbf{B} and \textbf{C} at time $t$ respectively, and let $S_a(0)$, $S_b(0)$ and $S_c(0)$ denote their initial states.
\begin{figure}[!t]
	\centering
		\includegraphics[width=3.5in]{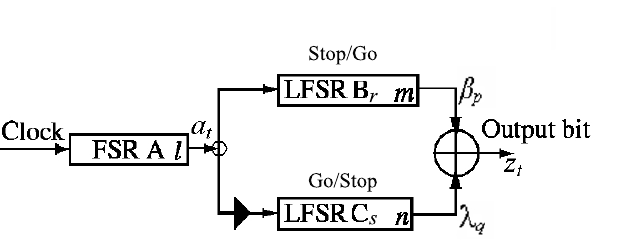}
	\caption{ASG model for the ASG$(r,s)$}
	\label{fig2}
\end{figure}
As the finite state machine is linear, the state transition can be described by a matrix which is the companion matrix for an LFSR. We refer to the transition matrix of registers \textbf{B} and \textbf{C} by $T_b$ and $T_c$ and we suppose that the matrixes $T_b$ and $T_c$ are known in the rest of the paper. So, we have:
\begin{equation}
	S_b(t) = S_b(t-1)T_b= S_b(0)T_b^t,
	\label{E1}	
\end{equation}
\begin{equation}
	S_c(t) = S_c(t-1)T_c= S_c(0)T_c^t.
	\label{E2}	
\end{equation}
Suppose that $z_t = b_i \oplus c_j$, so we have:
\begin{equation}
	z_{t+1} = (b_{i+r} \oplus c_j)a_t \oplus (b_i \oplus c_{j+s})(a_t \oplus 1).
	\label{E3}	
\end{equation}
Suppose that the first output bits of registers \textbf{B} and \textbf{C} are denoted by $b_0$ and $c_0$. It is clear that only the bits in positions $i=pr$ and $j=qs$ are chosen from the regular output sequence of registers \textbf{B} and \textbf{C} respectively and other bits are discarded. In other words, the keystream output sequence ($z_t$) is constructed by a combination of two $r$-decimated and $s$-decimated sequences derived from the regular output sequence of \textbf{B} and \textbf{C}. We refer to these irregular sequences by $\beta$ and $\lambda$ respectively. So, we have; $\beta = \beta_0, \beta_1 , ..., \beta_t = b_0, b_{1r} , ..., b_{tr}$, such that $\beta_t=b_{tr}$, for all $t\geq 0$ and $\lambda=\lambda_0, \lambda_1 , ..., \lambda_t = c_0, c_{1s} , ..., c_{ts}$, such that $\lambda_t=c_{ts}$, for all $t\geq 0$.

The constructor Kanso \cite{R[1]} recommended that each register \textbf{B} and \textbf{C} should be an LFSR with output being an $m$-sequence. According to the following well known theorem from \cite{R[14]}, both $\beta$ and $\lambda$ are $m$-sequences as well.
\newtheorem{theorem}{Theorem}
\begin{theorem}
Let $b$ be a binary maximum-length sequence ($m$-sequence) with periodicity ($2^m-1$). Let $\beta$ be a sequence obtained by sampling every $r^{th}$ bit of $b$, starting with the first bit of $b$. Then $\beta$ is again a $m$-sequence with the same period, if and only if $\gcd(r, 2^m-1)=1$.
\end{theorem}

This means that we can model the clock-controlled LFSRs \textbf{B} and \textbf{C}, by new regular LFSRs,  $\textbf{B}_r$ and $\textbf{C}_s$, with transition matrixes $T_\beta$ and $T_\lambda$ and regular output sequences $\beta = \beta_0, \beta_1 , ..., \beta_t$ and $\lambda=\lambda_0, \lambda_1 , ..., \lambda_t$ respectively. In other words, the sequences $\beta$ and $\lambda$ can be regenerated by the same length registers but different feedback polynomials. For their internal states, we have:
\begin{equation}
	S_\beta(t)=S_\beta(t-1)T_\beta =S_\beta(0)T_\beta^t,
	\label{E4}	
\end{equation}
\begin{equation}
	S_\lambda(t)=S_\lambda(t-1)T_\lambda=S_\lambda(0)T_\lambda^t.
	\label{E5}	
\end{equation}
If $E_b$ and $E_c$ denote the vectors which choose the last bit of registers \textbf{B} and \textbf{C}'s internal states as an output bit, we can write that:
\begin{equation}
	\beta_t=S_\beta(t)E_b = S_\beta(0)T_\beta^tE_b,
	\label{E6}	
\end{equation}
\begin{equation}	\lambda_t=S_\lambda(t)E_c = S_\lambda(0)T_\lambda^tE_c.
	\label{E7}	
\end{equation}
Suppose that $i=pr$ and $j=qs$, so we can rewrite (\ref{E3}) as follow:
\begin{equation}
	z_t = b_i\oplus c_j = \beta_p \oplus \lambda_q,\nonumber
\end{equation}
\begin{equation}
z_{t+1} = (\beta_{p+1} \oplus \lambda_q)a_t \oplus (\beta_p \oplus \lambda_{q+1})(a_t \oplus 1).
	\label{E9}	
\end{equation}
It can be recognized easily that (\ref{E9}) describes an original ASG whose output ($z_t$) is composed of $\beta$ and $\lambda$ under the control of $a_t$. So, we can model the ASG$(r,s)$ to the original ASG described in (\ref{E9}) which is illustrated in fig. \ref{fig2}. In the next section, we will use this model and algebraic techniques to attack the ASG$(r,s)$. 

Several attacks have been proposed on the original ASG in the literature, but most of them do not affect the security of the ASG$(r,s)$. Our idea can be applied to the original ASG, but it is not better than the previous attacks in contrast to the ASG$(r,s)$. 

Table~\ref{T1} shows the complexity of the previous attacks and our attack on the original ASG. In table~\ref{T1}, the first column shows the name of the previous attacks against the original ASG and the second column shows the \textbf{M}inimum \textbf{K}eystream \textbf{L}ength \textbf{R}equirement (MKLR). The third column shows the total complexity and the last column shows the complexity of the attack in the case when $l=m=n=64$. In table~\ref{T1} and table~\ref{T2}, $L$ and $M$ is equal to ($l+m+n$) and $\max\left\{m, n\right\}$ respectively, and also we have $\Gamma=1-1/(0.19m+3.1)$.

We can see easily from table~\ref{T1} that the Johansson's reduced complexity attack \cite{R[20]} is the best existing attack on the original ASG so far. For this reason, we briefly describe this attack and try to apply it to the ASG$(r,s)$. In the Johansson's reduced complexity attacks, the adversary waits for a segment of $M$ consecutive zeros (or ones) in the output sequence of the ASG. If $m\leq n$, then the adversary assumes that exactly $M/2$ of them are from LFSR \textbf{B}. This is true with probability:
\begin{equation}
	{M \choose M/2}2^{-M}.
	\label{E10}	
\end{equation}
The remaining ($m-M/2$) bits of LFSR \textbf{B} are found by exhaustive search. The optimal complexity of this attack on the original ASG is $O(m^22^{2m/3})$.

This attack can not be applied to the ASG$(r,s)$, because its main assumption, that exactly $M/2$ bits of the $M$-bits output segment comes from LFSR \textbf{B}'s initial state, is only true when the output is composed of the two Stop/Go Generators' output. But in case of ASG$(r,s)$, the values of $r$ and $s$ can be very large numbers. So, the main assumption to apply the Johansson's attack does not hold for the ASG$(r,s)$ in general. Therefore, we have to apply this attack to our ASG model for the ASG$(r,s)$, but it is not possible. Because the Johansson's attack needs to know the feedback polynomials of the generator registers, $\textbf{B}_r$ and $\textbf{C}_s$, but they are unknown in our ASG model. So, we have to search the $r$ and $s$ values to apply this attack. We can search these values in $\Phi$ steps and apply Johansson's attack for each value of the $r$ and $s$. The optimal complexity of this attack is $O(\Phi m^22^{2m/3})=O(m^22^{8m/3})$. In the next section, our attack on the ASG$(r,s)$ will be explained and compared to this attack in table~\ref{T2}.
\begin{table}[t]
	\caption{The Complexity Of Previous Attacks Against The Original ASG}
  \label{T1}
	\renewcommand{\arraystretch}{1.1}
	\centering
	\begin{tabular}{| >{\centering\arraybackslash}m{2.1cm} | >{\centering\arraybackslash}m{1.3cm} | >{\centering\arraybackslash}m{2.4cm}  | 	>{\centering\arraybackslash}m{1.1cm} |}
	
			\hline  Attack & MKLR  & Complexity &$l=m=n=64$\\
			
			\hline 	Edit Distance Correlation \cite{R[16], R[17], R[18]}& $O(m+n)$&$O((m+n)2^{m+n})$&$2^{135}$ \\
			
			\hline	Clock Control Guessing Attack\cite{R[22]}&$l+m+n$&$O(L^32^{L/2})$&$2^{118.8}$\\
			
			\hline	Algebraic Attack \cite{R[13]}&$O(m + n)$&$O((m^3+n^3)2^l)$&$2^{83}$\\
			
			\hline	Edit Probability Correlation Attack \cite{R[19]}&$O(m + n)$&$O(M^22^M)$&$2^{76}$\\
			
			\hline	Khazaei's Reduced Complexity Attack \cite{R[21]}&$2m$&$O(m^22^{\Gamma m})$&$2^{71.8}$\\
			
			\hline	Improved Edit Distance Correlation \cite{R[32]}&$O(M)$&$O(M2^M)$&$2^{70}$\\
			
			\hline	Linear Consistency Attack \cite{R[15]}&$ $&$O(\min(m,n)2^l)$&$2^{70}$\\
			
			\hline	Johansson's Reduced Complexity Attacks \cite{R[20]}&$O(2^{2m/3})$&$O(m^22^{2m/3})$&$2^{54.7}$\\
			
			\hline	Our Algebraic Attack &$3(m+n)$&$O((m^2+n^2)2^{l+1})$&$2^{78}$\\
			\hline
	\end{tabular}
\end{table}

\section{OUR ALGEBRAIC ATTACK ON THE ASG$(r,s)$}
The goal of an attack on the stream cipher is to recover the secret key or to predict and reproduce the rest of the keystream to recover the rest of the cipher text. In \cite{R[13]} an algebraic attack approach to a family of irregularly clock-controlled LFSR based systems is presented. The complexity of this attach on the original ASG structure is $O((m^3+n^3)2^l)$. But, its complexity on the ASG$(r,s)$ is approximately $O((m^3+n^3)2^{l+m+n-2})$. We make use of the same idea to attack the ASG$(r,s)$ but we have improved its complexity significantly. If we XOR $z_t$ by $z_{t+1}$ from (\ref{E9}), we have:
\begin{equation}
	z_t \oplus z_{t+1} {=} \beta_p \oplus \lambda_q \oplus ( \beta_{p+1} \oplus  \lambda_q)a_t \oplus  ( \beta_p \oplus \lambda_{q+1})(1 \oplus a_t).
	\label{E11}
\end{equation}
Now, if we multiply both sides of (\ref{E11}) by $a_t$, we have:
\begin{equation}
		(z_t \oplus z_{t+1})(a_t)=( \beta_p \oplus \beta_{p+1} )(a_t),
	\label{E12}
\end{equation}
and if we multiply both sides of (\ref{E11}) by $(1 \oplus a_t)$, we obtain:
\begin{equation}
		(z_t \oplus z_{t+1})(1 \oplus a_t)=( \lambda_q \oplus \lambda_{q+1} )(1 \oplus a_t).
	\label{E13}
\end{equation}
From (\ref{E12}) and (\ref{E13}) we conclude that:
\begin{equation}		
		if ~~~a_t=\begin{cases}
	            1 & \beta_{p+1} = \beta_p \oplus z_t \oplus z_{t+1}\\
	            0 & \lambda_{q+1} = \lambda_q \oplus z_t \oplus z_{t+1}
	  					\end{cases}.
	  					\label{E14}
\end{equation}
So, if we know the value of $a_t$, $\beta_p$ and $\lambda_q$, we can find $\beta_{p+1}$ and $\lambda_{q+1}$. Note that $z_t$ and $z_{t+1}$ belong to the known output sequence of the ASG$(r,s)$.

In our attack, we search over all possible values for the initial state of register \textbf{A} and produce the sequence $a= a_0, a_1, ..., a_t$. Then, we guess the value of $\beta_0$ and calculate $\lambda_0=z_0 \oplus \beta_0$. Now, by (\ref{E14}) we can find the bits $\beta_p$ for $p \geq 1$ and $\lambda_q$ for $q \geq 1$ as much as needed.

Using the Berlekamp-Massey algorithm and $2m$ bits of $\beta$ and $2n$ bits of $\lambda$, we can find the feedback polynomials and the initial states of the generator registers, $\textbf{B}_r$ and $\textbf{C}_s$, that can directly produce the sequences $\beta$ and $\lambda$ regularly. Then, by the rest of the output sequence we can test our guesses for the value of $\beta_0$ and the initial state of register \textbf{A}. As the complexity of Berlekamp-Massey algorithm is $O(n^2)$ for a sequence of length $n$, the complexity of this part of our attack is equal to $O((m^2+n^2)2^{l+1})$. 

Now, we have to find the value of parameters $r$ and $s$ and the initial states of LFSR \textbf{B}, $S_b(0)$, and \textbf{C}, $S_c(0)$. We first have to represent $b_{rt}$ and $b_t$ by the \emph{Trace Function}. The trace function, $Tr_m(x)$, is a mapping from the finite field $GF(2^m)$ to $GF(2)$ defined by 
\[ Tr_m(x)=\sum_{i=0}^{m-1} x^{2^i}. \] 
Any $m$-sequence $\{b_t\}$ of period $2^m-1$ with characteristic polynomial which is the minimal polynomial of a primitive element $\alpha$ (of order $2^m-1$) in $GF(2^m)$ can be represented by the trace function as $b_t  = Tr_m(u \alpha^t)$.
Every nonzero element $u \in GF(2^m)$ corresponds to a cyclic shift of $\{b_t\}$. In our case the situation is that we know $T_b$ and have found $T_{\beta}$ and we want to find $r$ and $s$. To find $r$ we know already $\alpha$ and $b_{rt}$ for $\left\{t=0, 1, 2, ...\right\}$ as well as the relations (\ref{trace:bt}) and (\ref{trace:b-rt}).
\begin{equation}
b_t  = Tr_m(u \alpha^t),
\label{trace:bt}
\end {equation}
\begin{equation}
b_{rt}  = Tr_m(u \alpha^{rt}) = Tr_m(u \gamma^t),
\label{trace:b-rt}
\end {equation}
where $(\gamma=\alpha^r)$. We want to find $u$ which is part of the key since it determines $\{b_t\}$. First we guess a possible value for $r$ and compute $\gamma=\alpha^r$. Then we construct an equation system by (\ref{trace:b-rt}) for $\{t=0, 1, 2, ..., m-1\}$. This is an equation system in the $m$ unknowns $u$, $u^2$, ..., $u^{2^{m-1}}$. The system has full rank due to the special form of the coefficient matrix and can therefore be solved in complexity $O(m^3)$. If the solution of equation system, $u$, can regenerate correctly the sequence $b_{rt}$ by using (\ref{trace:b-rt}) for $\{t= m, m+1, ...\}$ for sufficiently many bits, our guess for $r$ is correct. In other case, we have to repeat this process with new possible value for $r$. Then $u$ is found and we can generate $b_t$ by using (\ref{trace:bt}) for $\{t=0, 1, 2, ..., m-1\}$ which is the initial state of LFSR \textbf{B}. Similarly we can find the initial state in the other LFSR \textbf{C}. The complexity of this part is equal to $O(\Phi_1 m^3 + \Phi_2 n^3)=O(m^32^{m-1} + n^32^{n-1})$. Therefore, the total complexity of our attack is equal to:
\setlength{\arraycolsep}{0.0em}
	\begin{equation}
		\begin{split}
			C = O((m^2+n^2)2^{l+1}+ m^32^{m-1} + n^32^{n-1}).
			\label{E17}	
		\end{split}
	\end{equation}
\setlength{\arraycolsep}{5pt}
Table~\ref{T2} shows the complexity of the previous attacks and our attack on the ASG$(r,s)$ to compare their efficiencies. In case of $l=m=n=64$, the best previous attack needs $2^{153.5}$ steps to break the ASG$(r,s)$, but our attack is significantly better and it can find the secret key only by $2^{82}$ steps. This difference comes from our idea to find the values of $r$ and $s$.
\begin{table}[t]
	\caption{Comparison Of Our Attack On The ASG($r,s$) With Other Known Attacks}
  \label{T2}
	\renewcommand{\arraystretch}{1.1}
	\centering
	\begin{tabular}{| >{\centering\arraybackslash}m{2.1cm} | >{\centering\arraybackslash}m{1.3cm} | >{\centering\arraybackslash}m{2.4cm}  | 	>{\centering\arraybackslash}m{1.1cm} |}
	
			\hline  Attack & MKLR  & Complexity &$l=m=n=64$\\
			
			\hline	Clock Control Guessing Attack \cite{R[22]} & $l+m+n$ & $O(L^32^{\frac{L+2m+2n-4}{2}})$&$2^{566}$\\
			
			\hline 	Edit Distance Correlation \cite{R[16], R[17], R[18]}& $O(m+n)$&$O((m+n)2^{2(m+n)-2})$&$2^{261}$ \\
						
			\hline	Algebraic Attack \cite{R[13]} &$O(m+n)$&$O((m^3+n^3)2^{L-2})$&$2^{209}$\\
			
			\hline	Edit Probability Correlation Attack \cite{R[19]} &$O(m+n)$&$O(M^22^{M+m+n-2})$&$2^{202}$\\
			
			\hline	Improved Edit Distance Correlation \cite{R[32]} &$O(M)$&$O(M2^{M+m+n-2})$&$2^{196}$\\
			
			\hline	Linear Consistency Attack \cite{R[15]}&$ $&$O(\min(m,n)2^{3l-2})$&$2^{196}$\\
			
			\hline	Khazaei's Reduced Complexity Attack \cite{R[21]} &$2m$&$O(m^22^{(\Gamma +2)(m-2)})$&$2^{167.5}$\\
									
			\hline	Johansson's Reduced Complexity Attacks \cite{R[20]} &$O(2^{2m/3})$&$O(m^22^{(8m/3)-2})$&$2^{153.5}$\\
			
			\hline	Our Algebraic Attack &$3(m+n)$&$O((m^2+n^2)2^{l+1}+m^32^{m-1}~+~n^32^{n-1})$&$2^{82}$\\
			\hline
	\end{tabular}
\end{table}
In the previous attacks, the adversary has to guess the values of $r$ and $s$ by exhaustive search, and for each guess, the attack must be applied to the algorithm. But, in our idea, we do not need to know the values of $r$ and $s$ to apply our attack and we find these values independent of the exhaustive search over the initial state of register \textbf{A}.
 
\section{Conclusion}
In this paper, we present an ASG model for the ASG$(r,s)$ and also we present a new algebraic attack against the ASG$(r,s)$. The designer of the ASG$(r,s)$ claims that this structure is more secure than the original ASG, but we show that its security is not more than the original ASG. Our attack can find the secret key of ASG$(r,s)$ by using of $3(m+n)$ bits of the output keystream with $O((m^2+n^2)2^{l+1}+ m^32^{m-1} + n^32^{n-1})$ computational complexity. 

As far as we know, there is no efficient attack against the ASG$(r,s)$ so far. The complexity of previous attacks is much higher than the complexity of our attack. In case of $l=m=n=64$, the best previous attack needs $2^{153.5}$ steps to break the ASG$(r,s)$, but our attack can find the secret key only by $2^{82}$ steps.
Our attack can be applied to the original ASG structure. Its complexity is comparable to the best known attacks but our attack does not need to know the characteristic polynomial of generating registers. Applying our idea to other clock-controlled structures is a subject for future research.

\end{document}